\documentclass[%
aps,%
12pt,%
final,%
notitlepage,%
oneside,%
onecolumn,%
nobibnotes,%
nofootinbib,%
superscriptaddress,%
noshowpacs,%
centertags]%
{revtex4}
\usepackage[koi8-r]{inputenc}
\usepackage[english,russian]{babel}
\usepackage[T2A,T1]{fontenc}
\usepackage{epsfig}
\usepackage[centerlast,normalsize]{caption2}

\def\ktf {$k_t$-factorization }

\def\a {\epsilon}
\def\g {\gamma}

\def\F  {{\cal F}}
\def\p  {{\cal P}}
\def\J {$J/\psi$ }

\def\U {$\Upsilon$ }
\def\u { \Upsilon  }

\def\qq {$Q\bar{Q}$ }
\def\cpc#1#2#3  {{Computer\ Phys.\ Comm.\ }  {\bf#1}, #2 (#3)}
\def\err#1#2#3  {{\it Erratum }              {\bf#1}, #2 (#3)}
\def\epjc#1#2#3 {{Eur. Phys. J. C }          {\bf#1}, #2 (#3)}
\def\dum#1#2#3  {{~}                         {\bf#1}, #2 (#3)}
\def\ib#1#2#3   {{\it ibid. }                {\bf#1}, #2 (#3)}
\def\jcp#1#2#3  {{J.\ Comp.\ Phys.\ }        {\bf#1}, #2 (#3)}
\def\jhep#1#2#3 {{JHEP }                     {\bf#1}, #2 (#3)}
\def\ijmp#1#2#3 {{Int.\ J.\ Mod.\ Phys.\ }   {\bf#1}, #2 (#3)}
\def\jpg#1#2#3  {{J.\ Phys.\ G }             {\bf#1}, #2 (#3)}
\def\mpl#1#2#3  {{Mod.\ Phys.\ Lett.\ }      {\bf#1}, #2 (#3)}
\def\ncim#1#2#3 {{Nuovo Cimento }            {\bf#1}, #2 (#3)}
\def\np#1#2#3   {{Nucl.\ Phys.\ }            {\bf#1}, #2 (#3)}
\def\npb#1#2#3  {{Nucl.\ Phys.\ B}           {\bf#1}, #2 (#3)}
\def\pan#1#2#3  {{Phys.\ At.\ Nuclei }       {\bf#1}, #2 (#3)}
\def\plb#1#2#3  {{Phys.\ Lett.\ B }          {\bf#1}, #2 (#3)}
\def\prep#1#2#3 {{Phys.\ Rep.\ }             {\bf#1}, #2 (#3)}
\def\prd#1#2#3  {{Phys.\ Rev.\ D }           {\bf#1}, #2 (#3)}
\def\prl#1#2#3  {{Phys.\ Rev.\ Lett.\ }      {\bf#1}, #2 (#3)}
\def\ptp#1#2#3  {{Prog.\ Theor.\ Phys.\ }    {\bf#1}, #2 (#3)}
\def\ppnp#1#2#3 {{Prog.\ Part.\ Nucl.\ Phys.\ } {\bf#1}, #2 (#3)}
\def\ps#1#2#3   {{Physica Scripta }          {\bf#1}, #2 (#3)}
\def\rmp#1#2#3  {{Rev.\ Mod.\ Phys.\ }       {\bf#1}, #2 (#3)}
\def\rpp#1#2#3  {{Rep.\ Prog.\ Phys.\ }      {\bf#1}, #2 (#3)}
\def\sa#1#2#3   {{Sci. Acta}                 {\bf#1}, #2 (#3)}
\def\sjnp#1#2#3 {{Sov.\ J.\ Nucl.\ Phys.\ }  {\bf#1}, #2 (#3)}
\def\spj#1#2#3  {{Sov.\ Phys.\ JETP }        {\bf#1}, #2 (#3)}
\def\spjl#1#2#3 {{Sov.\ JETP Lett.\ }        {\bf#1}, #2 (#3)}
\def\spu#1#2#3  {{Sov.\ Phys.-Usp.\ }        {\bf#1}, #2 (#3)}
\def\yaf#1#2#3  {{Yad.\ Fiz.\ }              {\bf#1}, #2 (#3)}
\def\zp#1#2#3   {{Zeit.\ Phys.\ }            {\bf#1}, #2 (#3)}
\def\zpc#1#2#3  {{Z.\ Phys.\ C }             {\bf#1}, #2 (#3)}
\def\etal {{\it et al. }}
\begin{document}

\title{Upsilonium polarization as a touchstone\\
     in understanding the parton dynamics in QCD}
\author{C.\ P.\ Baranov}
\email{baranov@sci.lebedev.ru}
\affiliation{P.N.Lebedev Institute of Physics,
             Lenin avenue 53, Moscow 119991, Russia}
\author{N.\ P.\ Zotov}
\email{zotov@theory.sinp.msu.ru}
\affiliation{Institute of Nuclear Physics, Moscow State University,
             Lenin Hills, Moscow 119991, Russia}
\date{\today}
\begin{abstract} 
In the framework of the \ktf approach, the production of \U mesons 
at the Fermilab Tevatron and CERN LHC is considered,
and the predictions on the spin alignment parameter $\alpha$ are
presented. We argue that measuring the polarization of quarkonium
states can serve as a crucial test discriminating two competing
theoretical approaches to parton dynamics in QCD.
\end{abstract}
\maketitle

\section{INTRODUCTION}

Nowadays, the production of heavy quarkonium states at high energies 
is under intense theoretical and experimental study \cite{ref1,ref2}.
The production mechanism involves the physics of both short and long
distances, and so, appeals to both perturbative and nonperturbative 
methods of QCD. The creation of a heavy quark pair \qq proceeds via 
the photon-gluon or gluon-gluon fusion (respectively, in $ep$ and $pp$ 
collisions) referring to small distances of the order of $1/(2m_Q)$, 
while the formation of the colorless final state refers to longer 
distances of the order of $1/[m_Q\,\alpha_s(m_Q)]$. These distances 
are longer than the distances typical for hard interaction but are yet 
shorter than the ones responsible for hadronization (or confinement).
Consequently, the production of heavy quarkonium states is under 
control of perturbative QCD but, on the other hand, is succeeded by 
nonperturbative emission of soft gluons. This feature gives rise to 
two competing theoretical approaches known in the literature as the 
color-singlet \cite{BaiBer,GubKra} and color-octet \cite{ChoLei} models.
According to the color-singlet approach, the formation of a colorless
final state takes place already at the level of the hard partonic 
subprocess (which includes the emission of hard gluons when necessary).
In the color-octet model, also known as nonrelativistic QCD (NRQCD),
the formation of a meson starts from a color-octet \qq pair and proceeds 
via the emission of soft nonperturbative gluons.
The former model has a well defined applicability range and has already 
demonstrated its predictive power in describing the \J production at 
HERA, both in the collinear \cite{Kraem} and the \ktf \cite{j_dis} 
approaches. As it was shown in the analysis of recent ZEUS \cite{ZEUS}
data, there is no need in the color-octet contribution, neither in the 
collinear nor in the \ktf approach.

Originally, the color-octet model was introduced to overcome the 
discrepancy between the large \J production cross section measured in 
$pp$ interactions at the Tevatron \cite{CDF1,CDF2,CDF3} and the results 
of theoretical calculations based on the standard perturbative QCD 
technique. The problem was apparently solved by attributing the 
discrepancy to the hypothetical contributions from the intermediate 
color-octet states,
which must obey certain hierarchy in powers of the relative velicity of 
the quarks in a bound system. However, the numerical estimates of these 
contributions extracted from the analysis of Tevatron data are at odds 
with the HERA data, especially as far as the inelasticity parameter 
$z=E_{\psi}/E_{\g}$ is concerned \cite{KniZwi}.
In the \ktf approach, the values of the color-octet contributions 
obtained as fits of the Tevatron data appear to be substantially smaller 
than the ones in the collinear scheme, or even can be neglected at all
\cite{j_tev,Teryaev,Chao1,Vasin}.

In the present note we want to stress once again that measuring the 
polarizaton of quarkonium states produced at high energies may 
serve as 
an important and crucial test discriminating the different theoretical 
concepts.
The first attempts to solve the quarkonium polarization problem within
the \ktf approach were made in the pioneering work \cite{j_sha}
(see also \cite{Zotov}) for $ep$ 
collisions and in Refs. \cite{j_tev,Chao2} for $pp$ collisions.
It was emphasised that the off-shellness of the initial gluons, the
intrinsic feature of the \ktf approach, has an immediate consequence
(by analogy with longitudinal photons) in the longitudinal polarization
of the final state \J mesons. 
The theoretical predictions \cite{j_sha} have stimulated experimental 
investigation of \J spin alignment at the collider HERA. The first
results obtained by the collaborations H1 and ZEUS have been described
in Ref.\cite{j_dis}. These results have qualitatively confirmed the 
predictions on the dominance of longitudinal polarization.

The preliminary results on the \J polarization at the Tevatron obtained 
by the collaborations E537 \cite{E537} and CDF \cite{CDF4} also point to
logitudinal polarization with the average value of spin alignment 
parameter $\alpha\approx -0.2$ over the whole range of \J transverse 
momentum $p_T$. The collaboration D0 is currently analysing the data on 
the spin alignment of \U mesons.

In the NRQCD approach, the problem of quarkonium polarization remains 
unsolved \cite{ref1,ref2}. The gluon fragmentation mechanism leads 
to strong transverse polarization. Including the next-to-leading QCD 
corrections makes the transverse polarization even stronger. The only
way out is seen in increasing the fraction of unpolarized mesons by
attributing large contributions to the certain color-octet 
channels
\cite{Chao2}, which, however, violates the expected NRQCD 
hierarchy.
The role of the color-octet contributions taken into account in the
analysis of recent ZEUS data is obscure and does not lead to a
conclusive description of \J polarization parameters 
(see~\cite{ref2}).

\section{NUMERICAL RESULTS}

The goal of this paper is to derive theoretical predictions on the
polarization of \U mesons produced at the Fermilab Tevatron and CERN
LHC.
In the \ktf approach, the cross section of a physical process is
calculated as a convolution of the partonic cross section $\hat{\sigma}$
and the unintegrated parton distribustion ${\F}_g(x,k_{T}^2,\mu^2)$,
which depend on both the longitudinal momentum fraction $x$ 
and transverse momentum $k_{T}$:
\begin{equation}
  \sigma_{pp} =
  \int {\F}_g(x_1,k_{1T}^2,\mu^2)\,{\F}_g(x_2,k_{2T}^2,\mu^2)\,
  \hat{\sigma}_{gg}(x_1, x_2, k_{1T}^2, k_{2T}^2,...)
  \,dx_1\,dx_2\,dk_{1T}^2\,dk_{2T}^2.
\end{equation}
In accord with the \ktf prescriptions 
\cite{GLR83,Catani,Collins,BFKL},
the off-shell gluon spin density matrix is taken in the form
\begin{equation} \label{epsglu}
 \overline{\a_g^{\mu}\a_g^{*\nu}} =
  p_p^{\mu}p_p^{\nu}x_g^2/|k_{T}|^2 = k_{T}^\mu k_{T}^\nu/|k_{T}|^2.
\end{equation}
In all other respects, our calculations follow the standard Feynman
rules.

In order to estimate the degree of theoretical uncertainty connected
with the choice of unintegrated gluon density, we use two different
parametrizations, which are known to show the largest difference with
each other, namely, the ones proposed in Refs. \cite{GLR83,BFKL} 
and \cite{Bluem}.
In the first case \cite{GLR83}, the unintegrated gluon density is derived
from the ordinary (collinear) density $G(x,\mu^2)$ by differentiating it
with respect to $\mu^2$ and setting $\mu^2=k_T^2$.
Here we use the LO GRV set \cite{GRV98} as the input colinear density.
In the following, this will be referred to as dGRV parametrisation.
The other unintegrated gluon density \cite{Bluem} is obtained as a 
solution of leading 
order BFKL equation \cite{BFKL} in the double-logarithm approximation. 
Technically, it is calculated as the convolution of the ordinary gluon 
density with some universal weight factor. 
In the following, this will be referred to as JB parametrisation.

The production of \U mesons in $pp$ collisions can proceed via either 
direct gluon-gluon fusion or the production of $P$-wave states $\chi_b$
followed by their radiative decays $\chi_b{\to}\u{+}\g$.
The direct mechanism corresponds to the partonic subprocess
\begin{equation}\label{dir}
g+g\to\u+g
\end{equation}
which includes the emission of an additional hard gluon in the final 
state.
The production of $P$-wave mesons is given by
\begin{equation}\label{chi}
g+g\to\chi_b,
\end{equation}
and there is no emittion of any additional gluons.
As we have already mentioned above, we see no need in taking
the color-octet contributions into consideration.

The other essential parameters were taken as follows: 
the $b$-quark mass $m_b=m_\u/2=4.75$ GeV; 
the \U meson wave function $|\Psi_{\u}(0)|^2=0.4$ GeV$^3$ 
(known from the leptonic decay width $\Gamma_{l^+l^-}$ \cite{PDG});
the wave function of $P$-wave states $|\Psi_{\chi}'(0)|^2=0.12$ GeV$^5$
(taken from the potential model \cite{EicQui});
the radiative decay branchings
$Br(\chi_{b,J}{\to}\u\g)$ = 0.06, 0.35, 0.22 for $(J=0,1,2)$ \cite{PDG};
the renormalization and factorization scale $\mu^2 = m_{\u}^2+p_{T}^2$.

The polarization state of a vector meson is characterized by the spin
alignment parameter $\alpha$ which is defined as a function of any
kinematic variable as
\begin{equation}\label{alpha}
 \alpha(\p)=(d\sigma/d\p -3d\sigma_L/d\p)/(d\sigma/d\p +d\sigma_L/d\p),
\end{equation}
where $\sigma$ is the reaction cross section and $\sigma_L$ is the part 
of cross section corresponding to mesons with longitudinal polarization
(zero helicity state). The limiting values $\alpha=1$ and $\alpha=-1$
refer to the totally transverse and totally longitudinal polarizations.
We will be interested in the behavior of $\alpha$ as a function of the
\U transverse momentum: $\p\equiv |{\mathbf p}_{T}|$. 
The experimental definition of $\alpha$ is based on measuring the
angular distributions of the decay leptons
\begin{equation}\label{dgamma}
d\Gamma(\u{\to}\mu^+\mu^-)/d\cos\theta\sim 1+\alpha\cos^2\theta,
\end{equation}
where $\theta$ is the polar angle of the final state muon measured in 
the decaying meson rest frame.

The results of our calculations for the kinematic conditions of the
Tevatron and LHC are displayed in Figs. 1 and 2. In both cases, the
integration limits over rapidity were adjusted to the experimental
acceptances of CDF ($|y_\u|<0.6$) at the Tevatron and 
ATLAS ($|y_\u|<2.5$) at the LHC.
The upper panels show the predicted transverse momentum distributions.
Separately shown are the contributions from the direct (dashed lines) 
and $P$-wave decay (dotted lines) mechanisms. Note that, in 
spite of 
the suppression due to smaller values of $P$-wave finctions compared
to $S$-wave finctions, the dominant contribution comes from the
subprocess (\ref{chi}) rather than from (\ref{dir}).
The reason can be seen in the much smaller values of the final state
invariant masses, $m_\chi<<m_{\psi g}$. Our conclusion on the relative
size of two contributions is compatible with the preliminary estimates
obtained by the collaboration CDF \cite{CDF3}.
The $p_T$ shape of the individual contributions yet has not been
meaured experimentally. In the \ktf approach, this shape is 
determined 
by the unintegrated gluon density. The average $p_T$ is a bit 
lower in the process (\ref{dir}), because the total transverse
 momentum (equal
to that of the initial gluons) is shared between the two final state
particles; at the same time, the contribution from the matrix element
is nearly unimportant.

It is worth noting that the production of $\chi_b$ mesons can hardly
be described in a consistent way within the collinear factorization
scheme.
The leading order contribution coming from the subprocess (\ref{chi})
shows unphysical $\delta$-like $p_T$ spectrum. The usual lame excuses 
that the particles produced at zero $p_T$ disappear in the beam pipe 
and remain invisible do not work, because the decay products do have
nonzero $p_T$ and, undoubtedly, can be detected. At the same time,
the introduction of next-to-leading contributions (i.e., the processes
with extra gluons in the final state) causes the problem of infrared 
divergences.

The central panels in Figs. 1 and 2 show the behavior of the spin
alignment parameter $\alpha$ for \U mesons produced in the direct
subprocess (\ref{dir}).
The increase in the fraction of longitudinally polarised mesons is 
promptly connected with the increasing virtuality (and, consequently, 
the strenghtening longitudinal polarization) of the initial gluons.

As far as the decays of $P$-wave states are concerned, nothing is known
on the polarisation properties of these decays. If we assume that 
the
quark spin is conserved in radiative transitions, and the emission 
of
a photon only changes the quark orbital momentum (as it is known to be
true in the electric dipole transitions in atomic physics, 
$\Delta S=0$, $\Delta L=\pm 1$), then the predictions on $\alpha$
appear to be similar to those made for the direct channel (see lower
panels in Figs. 1 and 2, dotted curves). If, on the contrary, we
assume that the the transition $\chi_b{\to}\u + \g$ leads to 
complete
depolarization, then we arrive at a more moderate behavior of the 
parameter $\alpha$ (dash-dotted curves in Figs. 1 and 2). The overall 
polarization remains slightly longitudinal ($\alpha\simeq-0.2$) 
in the whole range of $p_T$ due to the 'direct' contribution.
A comparison between the data on \J and $\psi'$ polarization at the 
Tevatron \cite{CDF4} seems to give support to the depolarization
hypothesis. The difference between the \J and $\psi'$ polarization
data can be naturally explained by the presence of the depolarizing
contribution in the case of \J and the absense of this contribution
in the case of $\psi'$.

A state with purely direct production mechanism in the bottomonium 
family is the $\Upsilon(3S)$ meson. The calculations presented here 
are also valid for this state, except the lower total cross 
section 
(by an approximate factor of 1/3) because of the correspondingly lower
value of the wave function $|\Psi_{\u(3S)}(0)|^2=0.13$ GeV$^3$.
At the same time, the predictions on the spin alignment parameter
$\alpha$ remain intact (central panels in Figs. 1 and 2).

\section{CONCLUSIONS}

We have considered the production of \U mesons in high energy $pp$
collisions in the \ktf approach and derived predictions on the 
spin 
alignment parameter $\alpha(p_T)$. We point out that the predicted 
value of $\alpha(p_T)$ is typically negative in the whole range of 
$p_T$ and shows variations from $\alpha\simeq(-0.2)$ 
to $\alpha\simeq(-0.7)$ depending on the hypothesis assumed for the 
decays $\chi_b{\to}\u(1S){+}\g$.
At the LHC energies, the theoretical predictions possess less
sensitivity to the choice of unintegrated gluon distributions.
The purest probe is provided by the polarization of $\Upsilon(3S)$ 
mesons. In that case, the polarization is the strongest and the
predictions are free from uncertainties coming from the unknown
properties of $\chi_b$ decays.

We do not discuss the behavior of the parameter $\alpha(p_T)$ at
asymptotically large transverse momenta where the applicability of
the \ktf approach is questionable.

\acknowledgments

The authors thank V. Kuzmin for useful discussions on the experimental
situation at the D0 Collaboration.
N.Z. thanks P.F. Ermolov for support and DESY Directorate for support
and the hospitality. This research was also supported by the FASI of
Russian Federation (grant NS-8122.2006.2).

\newpage
\begin{figure}
\epsfig{figure=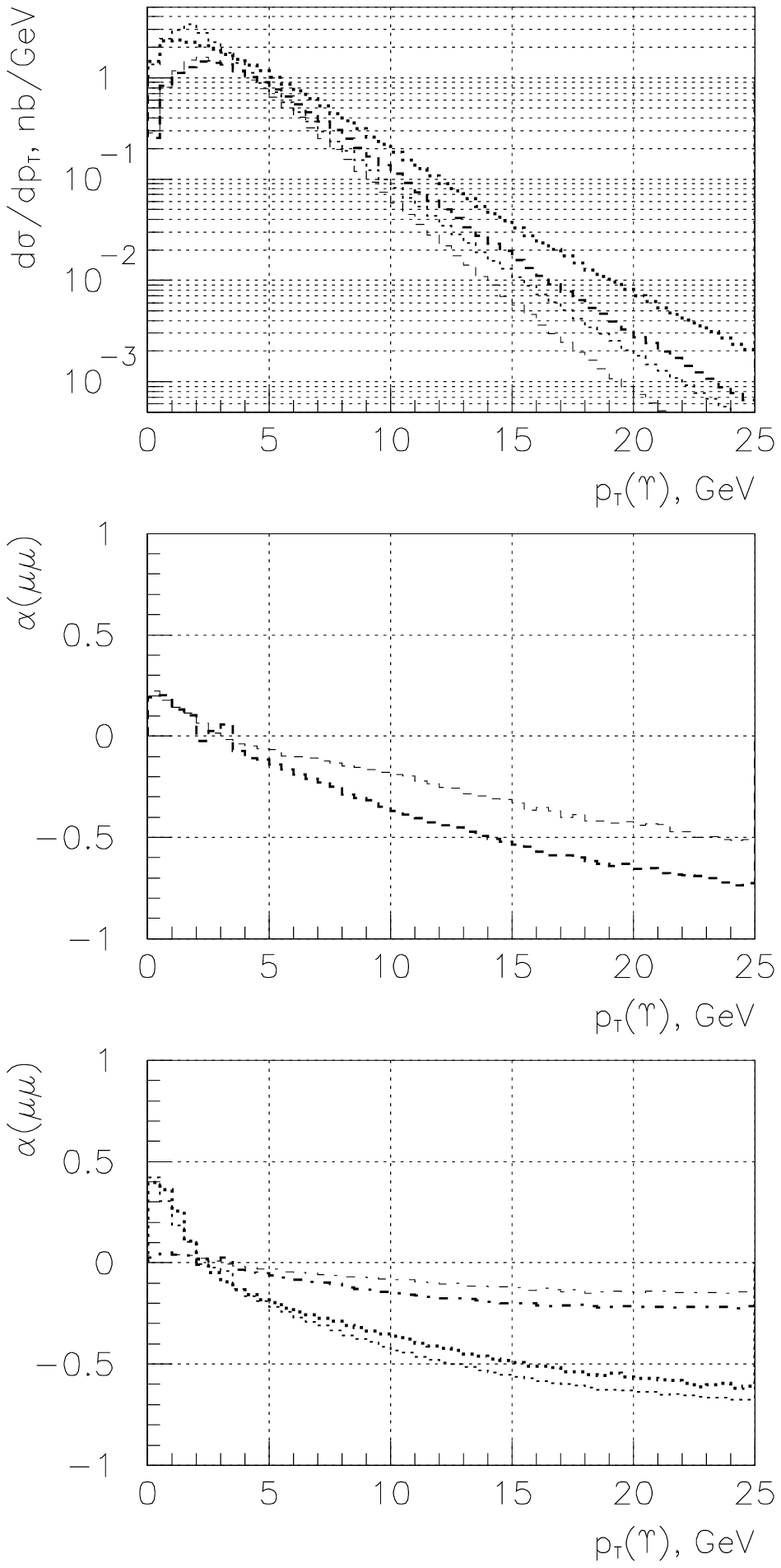, width = 10cm}
\caption{Predictions on the production of \U mesons at the Tevatron.
Thick lines, JB parametrization; thin lines, dGRV parametrization.
{\bf (a)} Transverse momentum distribution.
{\bf (b)} Spin alignment parameter $\alpha$ for the direct contribution.
{\bf (c)} Spin alignment parameter $\alpha$ with feed-down from $\chi_b$
 decays taken into account. 
 Dotted lines, the quark spin conservation hypothesis;
 dash-dotted lines, the full depolarization hypothesis.}
\label{fig1}
\end{figure}

\begin{figure}
\epsfig{figure=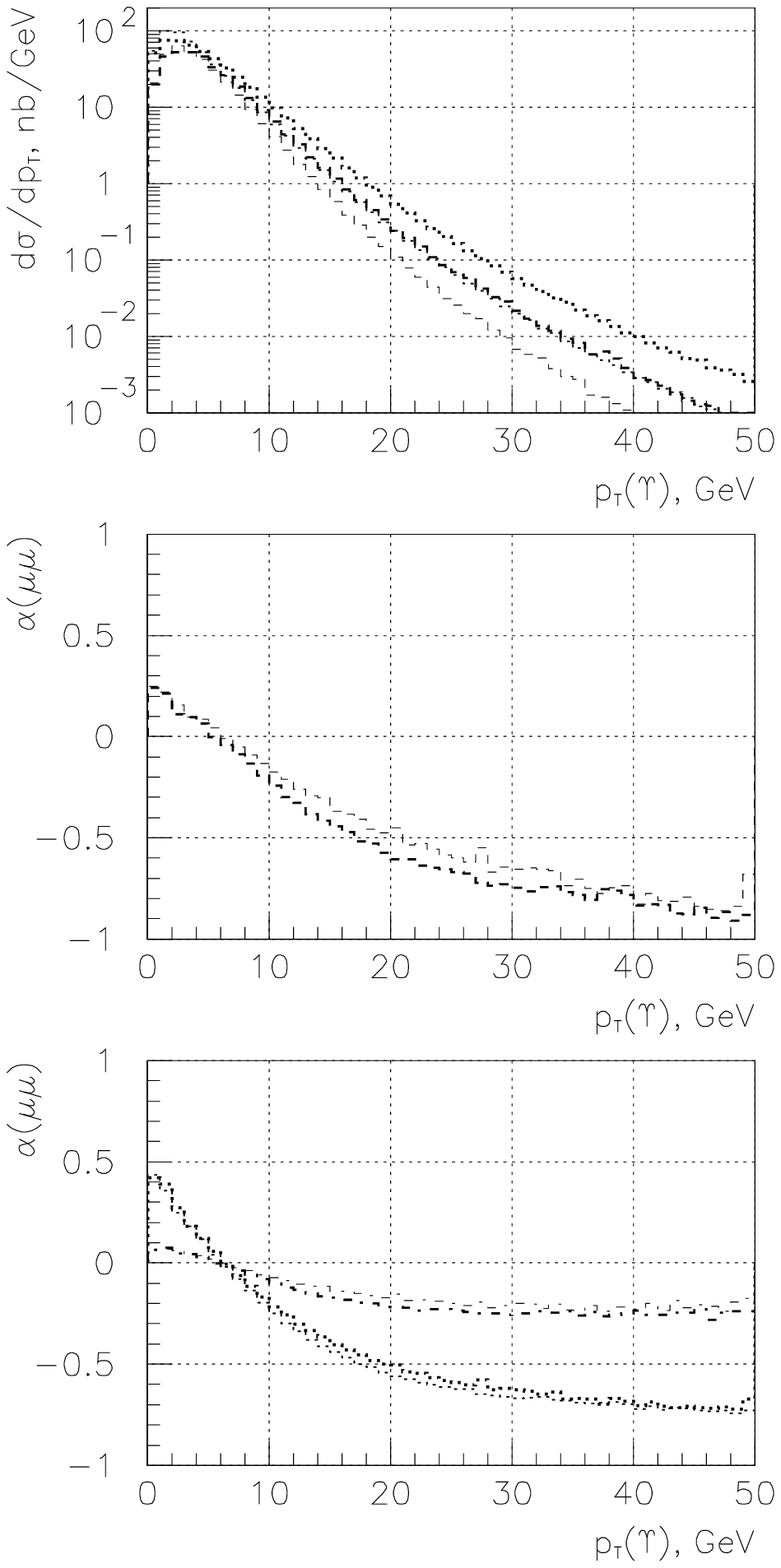, width = 10cm}
\caption{Same as Fig. 1, but for the LHC conditions.}
\label{fig2}
\end{figure}


\begin{thebibliography}{99}
\bibitem{ref1} M.\ Kr\"{a}mer,  \ppnp{47}{141}{2001} ;
J.~P.~Lansberg, \ijmp{A21}{3857}{2006} .
\bibitem{ref2} N.\ Brambila \etal, CERN-2005-005, hep-ph/0412158.
\bibitem{BaiBer} C.-H.\ Chang,                 \np{B172}{425}{1980} ;
                 R.\ Baier and R.\ R\"uckl,    \plb{102}{364}{1981} ;
                 E.\ L.\ Berger and D.\ Jones, \prd{23}{1521}{1981} .
\bibitem{GubKra} H.\ Krasemann,                \zpc{1}{189}{1979} ;
                 G.\ Guberina, J.\ K\"uhn, R.\ Peccei, and R.\ R\"uckl,
                                               \np{B174}{317}{1980} .
\bibitem{ChoLei} P.\ Cho and A.\ K.\ Leibovich,
                     \prd{53}{150}{1996} ;  \dum{53}{6203}{1996} .
\bibitem{Kraem}  M.\ Kr\"amer,              \np{B459}{3}{1996} .
\bibitem{j_dis} S.\ P.\ Baranov and N.\ P.\ Zotov, \jpg{29}{1395}{2003} ;
                A.\ V.\ Lipatov and N.\ P.\ Zotov, \epjc{27}{87}{2003} .
\bibitem{ZEUS}  ZEUS Collaboration, S.\ Chekanov \etal,
                                                   \epjc{44}{13}{2005} .
\bibitem{CDF1} CDF Collaboration, F.\ Abe \etal, \prl{69}{3704}{1992} ;
    \dum{71}{2537}{1993} ; \dum{75}{1451}{1995} ; \dum{79}{578}{1997} .
\bibitem{CDF2} CDF Collaboration, T.\ Affolder \etal,
                          \prl{84}{2094}{2000} ; \dum{85}{2886}{2000} .
\bibitem{CDF3} CDF Collaboration, F.\ Abe \etal, \prl{86}{3963}{2001} .
\bibitem{KniZwi} B.\ Kniehl, L.\ Zwirner,        \npb{621}{337}{2002} .
\bibitem{j_tev} S.\ P.\ Baranov,             \prd{66}{114003}{2002} .
\bibitem{Teryaev} Ph.\ H\"agler, R.\ Kirschner, A.\ Sh\"afer,
         L.\ Szymanowski, and O.\ V.\ Teryaev, \prd{63}{077501}{2001} ;
                                               \prl{86}{1446}{2001} .
\bibitem{Chao1} F.\ Yuan and K.-T.\ Chao,      \prd{63}{034006}{2001} .
\bibitem{Vasin} B.\ Kniehl, A.\ Saleev, D.\ Vasin,
                      \prd{73}{074022}{2006} ; \dum{74}{014024}{2006} .
\bibitem{j_sha} S.\ P.\ Baranov,               \plb{428}{377}{1998} .
\bibitem{Zotov} S.\ P.\ Baranov, A.\ V.\ Lipatov, N.\ P.\ Zotov,
         Proc. of the 9th Int. Workshop on DIS and QCD, Bologna, Italy
         (2001), p. 441; hep-ph/0106229.
\bibitem{Chao2} F.\ Yuan and K.-T.\ Chao,      \prl{87}{022002}{2001} ;
                                               \prd{63}{034017}{2001} .
\bibitem{E537} E537 Collaboration, C.\ Akerlof \etal,
                                           \prd{48}{5067}{1993} .
\bibitem{CDF4} CDF Collaboration, T.\ Affolder \etal,
                                           \prl{85}{2886}{2000} .
\bibitem{GLR83}
 L.\ V.\ Gribov, E.\ M.\ Levin, and M.\ G.\ Ryskin, \prep{100}{1}{1983} ;
                E.\ M.\ Levin and M.\ G.\ Ryskin, \prep{189}{267}{1990} .
\bibitem{Catani} S.\ Catani, M.\ Ciafaloni, and F.\ Hautmann,
                           \plb{242}{97}{1990} ; \np{B366}{135}{1991} .
\bibitem{Collins} J.\ C.\ Collins and R.\ K.\ Ellis, \np{B360}{3}{1991} .
\bibitem{BFKL} E.\ A.\ Kuraev, L.\ N.\ Lipatov, and V.\ S.\ Fadin,
                           \spj{45}{199}{1977} ;
               Ya.\ Balitsky and L.\ N.\ Lipatov, \sjnp{28}{822}{1978} .
\bibitem{Bluem} J.\ Bl\"umlein, \jpg{19}{1623}{1993} ; DESY 95-121 
(1995).
\bibitem{GRV98} M.\ Gl\"uck, E.\ Reya, and A.\ Vogt, 
\epjc{5}{461}{1998}.
\bibitem{PDG} Particle Data Group, W.-M.\ Yao \etal, \jpg{33}{1}{2006} .
\bibitem{EicQui} E.\ J.\ Eichten and C.\ Quigg, \prd{52}{1726}{1995} .
\end{thebibliography}
\end{document}